\documentclass{easychair}

\usepackage{doc}
\usepackage{multirow}
\usepackage{arydshln}
\usepackage{tcolorbox}
\usepackage{placeins}
\usepackage[colorinlistoftodos,prependcaption,textsize=tiny]{todonotes}
\usepackage{tikz}
\usepackage{enumitem}

\usetikzlibrary{shapes.geometric, arrows}
\tikzstyle{process} = [rectangle, minimum width=2.2cm, minimum height=1.7cm, text centered, draw=black]
\tikzstyle{arrow} = [thick,->,>=stealth]

\newcommand*\circled[1]{\raisebox{.5pt}{\textcircled{\raisebox{-.9pt} {#1}}}}

\title{Quality Control Methodology for Simulation Models\\of Computer Network Protocols}

\author{
    Vladimír Veselý\inst{1}
\and
    Jan Zavřel\inst{1}
}

\institute{
  Faculty of Information Technology, Brno University of Technology, Czech Republic\\
  \email{veselyv@fit.vutbr.cz, xzavre10@stud.fit.vutbr.cz}
}

\authorrunning{Veselý and Zavřel}

\titlerunning{Quality Control Methodology for Simulation Models of Computer Network Protocols}

\begin{document} 

\maketitle

\begin{abstract}
This paper summarizes know-how about modeling and simulation of computer networking protocols we contributed to the OMNeT++ community. We propose a methodology aiming to set a reliable ground truth for the quality of simulation models of networking protocols. We demonstrate the application of this methodology on our EIGRP source code pull-requested to the INET framework.
\end{abstract}

%
%

\section{Introduction}
\label{sec:intro}
The quality of simulation results depends on the accuracy of used models, namely, on how precisely models reflect the behavior of the real-world system. 
In computer science, developing a proper model and subsequent employment of this model in the running simulation is about finding a balance between complexity and effectiveness. 
Model implementation should be accompanied by reproducible proof to show how verification and validation (V\&V) was done. 
Thanks to that, other programmers can assess the accuracy of the model and its feasibility in the frame of different scenarios.

IEEE 1012 \cite{IEEE1012} standard defines V\&V  as processes, which determine whether developed products of a given activity conform to the requirements of that activity and whether the software satisfies its intended use and user needs. 
V\&V processes should provide an objective assessment of software products, namely, demonstrate software correctness, completeness, accuracy, consistency, and testability. 
IEEE standard, of course, tries to accommodate whatever type of software engineering effort, while we will apply its core principles in the field of computer modeling and simulation.

The domain of computer network protocols is our primary research and teaching interest. 
We are periodic contributors of simulation models for the OMNeT++ ecosystem. 
This paper focuses on our experience with the testing of developed models and their comparison with referential implementations. 
Our goal is to define and demonstrate the methodology that is helping us to produce precise simulation models.

This paper has the following structure. Section \ref{sec:sota} elaborates on inputs of the simulation model development process. Section \ref{sec:methodology} describes the proposed methodology on how to validate, verify and test simulation models; moreover, this section discusses certain challenges and introduces useful practices. Section \ref{sec:demo} demonstrates methodology on the use-case of EIGRP models for OMNeT++ and its framework INET. The paper is concluded in Section \ref{sec:conclusion}.

\section{State of the Art}
\label{sec:sota}
The following subsections present different ways of formal description of network protocols, highlight completely different approaches by their authors in their design and unravel general best practices of the V\&V process. We provide a short overview of the most commonly used referential implementations by the networking community at the end of this section.

\subsection{Protocol Definition}
Any layered model (e.g., ISO/OSI, TCP/IP, RINA \cite{RINA}) of computer network slices its functionality onto a set of collaborating protocols. 
Many computer protocols exist to handle different use-cases due to each protocol's unique functionality within the fixed scope (i.e., layer).
Basically, all computer networking protocols are defined by the rules of communication, which specify syntax, semantics, and timing of messages. 
These rules can be formally described either using deterministic finite-state machines (FSM) or temporal logic.

Theory and understanding of FSMs are less comprehensive and more accessible to programmers than temporal logic.
Therefore, computer network protocol modeling means conversion of protocol rules onto FSM, where 1) \textit{FSM states} represent internal communication statuses; and 2) \textit{FSM transitions} are a set of actions initiated by messages/events. 
FSM as a protocol model is then implemented using means (i.e., languages, built-in tools) available for the selected domain. 

It depends on the domain if means for behavior/state description and message syntax definition are different or the same. 
For instance in OMNeT++, we use C++ (\texttt{*.cc/*.h} files) to implement behavior, NED language (\texttt{*.ned} files) to layout composition, and message definitions (\texttt{*.msg} files) for syntax. In the case of network stack within the operating system (OS), most data-link, network and transport layer protocols (including their messages) are implemented via C/C++.
For message definition purposes, there even exist (semi)automated approaches that will help with the conversion of protocol specification into source code such as ASN.1\footnote{ASN.1 is codified in the standard available at https://www.itu.int/rec/T-REC-X.680-201508-I/en} or Google Protocol Buffer\footnote{Interface description language including code generator is available in the following GitHub repository https://github.com/protocolbuffers/protobuf }.

\subsection{Referential Implementations}
\label{sec:ref}
Standards – such as RFCs, or IEEE 802 series – outline all mandatory protocol rules so that different implementations are compatible with each other. 
Nevertheless, some standards allow or even encourage the flexibility of implementation. 
For instance:

\begin{itemize}
    \itemsep-0.2em
    \item IPv6 is using Extension Headers to add functionality that is not present in the Fixed Header. New Extension Headers handling unique use-cases can be created based on predefined rules and recommendations  \cite{IPv6} (section 4, pages 7 -- 25). Even the structure of already defined Extension Headers can be the subject of discussion, where different parties would implement it according to their needs (e.g., \cite{DnsInRA}).
    \item Similarly to that IS-IS, EIGRP \cite{EigrpRFC} (section 6.6, pages 51 -- 52), TCP \cite{TCP} (section 3.1, pages 17 -- 19), and others are often using the concept of type-length-value (TLV) fields to support easy updates of protocol functionality without the need to standardize a new version as is the case with OSPF for example.
\end{itemize}

The examples provided above show that the standard itself is not the ultimate source of protocol definition because actual implementations can also contribute to overall protocol functionality. 
Therefore, the V\&V process should include the comparison of the simulation model with another implementation of the selected protocol.	
Choosing the referential implementation has a significant impact because messages and results generated by this reference affect the development process of the simulation model. 
It would be wise to compare the simulation model with multiple referential implementations in order to obtain an objective and comprehensive evaluation. 
However, this may be in contradiction with the availability of resources (e.g., time, hardware) to the programmer. 
So if just a single referential implementation is used, let it be the one closest to the simulated phenomenon.

In our opinion, the best practice for referential implementation is to use an active network device (like a router or a switch) from a well-known vendor.
In that case, we have a chance to test multiple referential implementations on the same device thanks to generally available firmware updates. 
Moreover, employers of well-known vendors often co-create (or at least actively participate in discussions) during the standard development, which makes vendor implementations more trustworthy regarding compliance.

Here is a short non-exclusive list of referential implementations we have seen being used with respect to INET contributions:

\begin{itemize}
    \itemsep-0.2em
    \item \textbf{Cisco Packet Tracer} \cite{packetTracer}: 
    It is a tool associated with the Cisco NetAcad; it is used as a learning environment supporting the teaching of the curriculum of various computer networking classes.
    Diving deeper into details, Cisco Packet Tracer is a discrete event simulator containing models mimicking the behavior of selected products (mostly small office/home office routers, switches, access points, and security appliances) from the Cisco portfolio.
    Simulation models features have significant differences between different Cisco Packet Tracer versions.
    Based on version changelogs \cite{packetTracerFeatures}, community observations, and our insight, Cisco Packer Tracer embeds only limited equipment features.
    The majority of supported protocols are simplified or altered in behavior or message structure. 
    
    \item \textbf{Physical device}: 
    If possible, then the physical device with operating systems tailored for computer networking is the best referential implementation one can have. 
    However, specific protocols might be purposely constrained or not available to align with the business missions of such devices.
    On the one hand, open standard TRILL protocol is often omitted at the expense of similar yet proprietary protocols (e.g., FabricPath).
    On the other hand, originally proprietary protocol EIGRP was released to the public domain, but other vendors simply ignore it.
    Another cavity of employing physical devices is the cost because it may be expensive to build analogous topology to the simulated scenario.
    
    \item \textbf{GNS3} \cite{GNS3}, \textbf{EVE-ng} \cite{EVE-NG}: 
    GNS3 and EVE-ng are emulators of active network devices. 
    They can unpack and run binary firmware images (e.g., Cisco IOS files, QEMU images) of these devices in virtualized environments.
    This means that protocol implementation on the virtualized device is the same as on the real one. 
    GNS3 and EVE-ng employ the same virtualization (either QEMU \cite{QEMU} or Dynamips \cite{Dynamips}), but they differ only in the end-user interface; while GNS3 requires a dedicated application installed, EVE-ng wraps its functionality in the form of a web application. 
    Using both GNS3 or EVE-ng, the programmer may easily create complex topologies running tens even hundreds of nodes (the only limitation is available CPU and memory on the hosting computer) running reliable physical device implementation for nearly zero cost.

\end{itemize}

We consider Cisco Packet Tracer as an inferior choice for referential implementation due to its limitations.
The physical device is always the best option when it comes to accuracy to real-life protocol behavior.
However, it is bold to expect that other programmers who want to repeat the V\&V process would possess exactly the same number of devices with the same peripheral configurations and firmware versions.
We prefer to use the emulators mentioned above because their implementation is identical to physical devices.
Moreover, V\&V process reproduction is more effortless because both these tools offer convenient exports of topologies (including their running configuration and other state properties), which can be easily attached or referenced in the repositories of simulation models.

\section{Quality Control}
\label{sec:methodology}
This paper aims to define a structured V\&V process that any programmer may use as a cookbook for quality control of simulation models. 
This section briefly informs about various challenges which may be encountered during the development and testing phases. 
Then it describes each step of methodology based on all previously mentioned observations in this article.

\subsection{Challenges}
\label{subsec:challenges}
Here goes a curated and non-definitive list of challenges impacting development.
We provide our personal view on dealing with these challenges and welcome any counter-opinions.

\subsubsection{Level of Accuracy}
\label{subsubsec:accuracy}
A complete conversion of all protocol rules may lead to extremely complicated FSM (with many states/transitions and complex message variants).
The situation gets even more tricky with protocols that offload signalization or data transfer onto other protocols – e.g., IPsec ESP/AH cooperating with ISAKMP, RTCP/RTP cooperating with H.323/SIP.

There are approaches and complete tools capable of FSM minimization that would remove unnecessary states or transitions. 
However, this type of optimization was most probably already done by authors during the protocol design. 
The only other option for reducing complex FSM is to purposely omit a subset of protocol features.
This means a decrease in the accuracy of the simulation model. There is no rule of thumb for what to include and exclude because it depends on the programmer's goals.
Nevertheless, the following section might be helpful when questioning the desired level of accuracy of implemented computer network protocol.

\subsubsection{Application of Cryptography}
\label{subsubsec:cryptography}
Cryptography is being used in the protocol design to guarantee confidentiality (no one except sender and receiver can understand the message), integrity (any tempering with the message is recognized by the receiver), and authenticity (identity of single or both parties of the communication are guaranteed). 
The programmer needs to decide whether or not to include cryptography in the simulation model.

There are two main arguments for why to do it: 1)~the messages generated by the simulator would be the same as messages produced by referential simulation; thus 2)~it is the only way how to also support hardware in the loop (HIL) simulation. The following list represents counter-arguments why not add cryptography: 1)~there is no reason why to support cryptography primitives in deterministic simulation scenarios – it is a known fact before running the simulation whether confidentiality/integrity/authenticity is guaranteed or not between involved parties; 2)~the boilerplate of the simulation model source code tends to increase dramatically by adding external libraries handling cryptography (such as OpenSSL \cite{OpenSSL}); which leads to 3)~cryptography poses an overhead on resources (mainly CPU time and memory) when running the simulation, which means we need to wait longer for results or we could be even unable to simulate complex topologies with cryptography enabled.

Since we have never considered HIL simulations as the use-case for our models, we skip real-life implementation of the cryptography primitives. 
To achieve a comparable message structure, we replace relevant fields within the protocol message with an appropriate magic string. The magic string can be used as piggybacking mechanism for the receiver that the message “is encrypted/authenticated” or “with intact integrity”.

\subsubsection{Timing}
\label{subsubsec:timing}
Time is of the essence when comparing protocol messages' confluence between devices running referential implementation and simulation models. 
Referential implementation of the protocol runs in real-time, while the simulation is governed by a discrete event scheduler (for OMNeT++ see \cite{OmnetManual}, sections 4, 17 and 18). 

Due to the lack of global clocks, it is hard to measure durations, trigger actions, and control events between devices in real-time.
Therefore, it is absolutely mandatory to employ time synchronization protocols like NTP \cite{NTP} or PTP \cite{IEEE1558} for devices running referential implementation. 
NTP aims to achieve millisecond-level, while PTP guarantees up to nanosecond-level synchronization accuracy. 
Therefore, PTP is always a better choice; unfortunately, most active network devices support only NTP. If and only if the clocks are synchronized, the programmer is able to produce reliable baselines in the topologies running referential implementations.

Every discrete event simulator has built-in mechanisms to schedule events and trigger them during the simulation run. 
In the case of the INET framework, \texttt{ScenarioManager} \cite{ScenarioManager} offers setting up and control over simulation via scripted execution of events defined in the XML file (see for example Figure \ref{fig:scenariomanager}). 

Scheduling the event in the environment running referential implementation is more complicated (whether the device is real or virtualized).
It depends on the vendor, what scripting options are available. For instance, operating systems of many Cisco devices offer Embedded Event Manager (EEM) \cite{EEM} (the configuration snippet disabling specific interface at the given moment is depicted in Figure \ref{fig:applet}) or TCL \cite{TCL}. 
Instead of preloaded scripts, the programmer may use a more centralized approach (e.g., Expect \cite{Expect}, Ansible \cite{Ansible}).
Scripts are present on a single machine that would connect remotely (using SSH, Telnet, HTTP) to the device at the right moment.
Scripts would then be executed through a remote connection.
In the case of time-driven scripts, this approach offers more robustness since there is only a single clock controlling execution and better visibility for the scheduling of event batches.

\subsubsection{Control Plane Randomness}
\label{subsubsec:controlplane}

The control plane of the actual device runs many different processes, where each is responsible for a different protocol or functionality. 
The control plane dynamically switches between these processes based on OS resource schedulers.
This context switching introduces a degree of randomness, which impacts the reproducibility and baselines' readability.
Following symptoms relate to this challenge:

\begin{itemize}
    \itemsep-0.2em
    \item Stochastic delays are observed in the functionality of referential implementation when the control plane is preoccupied with another process;
    \item Consecutive protocol messages have non-standard gaps between each other due to the packet pacing. This jitter between messages is purposely introduced by the control plane either to avoid potential racing conditions between protocol instances or to guarantee stable bandwidth consumption.
\end{itemize}

Any simulation model cannot accurately abstract above mentioned symptoms.
Hence, they cannot be replicated inside the simulation.
This means that any comparison between referential implementation and its simulated variant should consider the factor of the control plane randomness.

\subsection{Methodology}
\label{subsec:method}

At least any V\&V process is a natural part of any software engineering effort.
We are sure that every programmer is conducting it in the form of source code testing.
In this section, we are not codifying anything new for the OMNeT++ community.
We just want to articulate quality control methodology related to the development of simulation models, including best practices based on the previous elaboration of our thoughts.

The proposed methodology consists of six consecutive phases depicted in the following diagram and described below:

\begin{figure}[!ht]
    \centering
    \scalebox{0.82}{
\begin{tikzpicture}[node distance=3cm]
\node (1) [process, align=center] {\circled{1} Choosing\\Referential\\Implementation};
\node (2) [process, right of=1, align=center] {\circled{2} Building of\\Testing\\Topology};
\node (3) [process, right of=2, align=center] {\circled{3} Baselines\\Production};
\node (4) [process, right of=3, align=center] {\circled{4} Comparison};
\node (5) [process, right of=4, align=center] {\circled{5} Revisiting the\\Implementation};
\node (6) [process, right of=5, align=center] {\circled{6} V\&V\\Reproduction\\Package};

\draw [arrow] (1) -- (2);
\draw [arrow] (2) -- (3);
\draw [arrow] (3) -- (4);
\draw [arrow] (4) -- (5);
\draw [arrow] (5) -- (6);
\draw [arrow] (5.south) |-  (9,-1.5) -- (4.south);
\draw [dashed,arrow] (9,-1.5) -- (6,-1.5) --  (3.south);
\end{tikzpicture}
}
\end{figure}
\FloatBarrier

\begin{enumerate}[label=\protect\circled{\arabic*}]
    \itemsep-0.2em
    \item \textbf{Choosing Referential Implementation}: The initial phase is about choosing the proper referential implementation compatible with expected simulation goals. As stated in section \ref{sec:ref}, we recommend either physical devices from a trusted vendor or a network emulator with a corresponding firmware image.
    
    \item \textbf{Building of Testing Topology}: Before employing a developed simulation model in the large-scale simulation, it is crucial to successfully conduct V\&V on the smallest possible topology, which would offer a good testing ground to assess normal behavior and treatment of edge cases. It is important to keep parameters (e.g., interface speeds, IP subnetting, number of adjacent devices, protocol-specific configuration) constant across real and simulated topology to maintain integrity.
    
    \item \textbf{Baselines Production}: We predominantly use the following three types of baselines produced by referential implementation: a) Syslog messages that include information about overall control-plane status; b) outputs of \texttt{show}/\texttt{debug} commands that describe in detail operation of referential implementation; c) PCAP files that contain computer traffic dumps. While Syslog and \texttt{show}/\texttt{debug} outputs are specific for a single device; PCAP files offer a view on protocol context between multiple devices since capture contains data from both directions (i.e., incoming/outgoing to/from selected interface). All above-mentioned baselines need to include time information which is important for subsequent comparison with the corresponding simulation scenario. The accuracy for all baseline types can be potentially enhanced up to nanosecond precision.
    
    \item \textbf{Comparison}: Whenever a developed simulation model is ready, it is being tested in preferably different simulation topologies and various scenarios. All of these test runs produce results (such as console message, trace, or PCAP) that are inputs for the comparison. The rest of this phase is about objective assessment between referential implementation and simulation model is made. We recognize two levels of such comparison: \textit{protocol level}, where we are focusing on the generated messages and their integrity (both syntactical and semantical); and \textit{abstract data structure level}, which focuses on states of abstract data structures used by the protocol, such as the routing table, interface table, CAM table, topology table for EIGRP, link-state database for OSPF, etc. Needless to say that a combination of both levels achieves the most objective evaluation. Thanks to the impact of discussed challenges, differences of various magnitude can be encountered during the comparison. Hence, it is important to conduct baselines production and comparison repeatedly. There is a very thin line between making the objective comparison of ground-truth baseline and simulated behavior, and subjectively choosing matching simulation results onto the corresponding baseline.
    
    \item \textbf{Revisiting the Implementation}: The simulation model can be modified, updated, or even completely redesigned depending on findings from the previous step. After all the changes are introduced to the implementation, it is wise to revert back to phases \circled{4} or \circled{5}, to verify the validity of changes. This process is repeated until the quality of the simulation model is sufficient (hopefully, the quality would even exceed original expectations).
    
    \item \textbf{V\&V Reproduction Package}: This phase is a memo that result reproduction is essential. Any contributed simulation model should be accompanied by materials (e.g., referential implementation version, baselines including PCAP/Syslog dumps, simulation trace files) that allow reproduction of resulting behavior as proclaimed by the author. Additional testing and V\&V done by the community have a chance to find new errors or unhandled cases that may further improve the quality of resulting simulation models.
\end{enumerate}

\section{Demonstration}
\label{sec:demo}
This section contains a demonstration of the V\&V methodology proposed in Section \ref{sec:methodology}. We decided to demonstrate this process on INET's EIGRP simulation model as available in INET 4.3 running in OMNeT++ 6.0 pre10. The simulation model is configured via a combination of \texttt{Ipv4NetworkConfigurator} for assigning IP addresses and EIGRP specific \texttt{.xml} file which has a deliberately similar structure to Cisco configuration.

\subsection{Choosing Referential Implementation}
Currently, the only viable choice for referential implementation of EIGRP is Cisco. Firstly, Cisco Systems is the author of this protocol. As such, there should be the smallest amount of discrepancies between the RFC and actual implementation caused by genuine errors and mistakes in the implementation. Secondly, there are very few other implementations that could be used as a reference. However, some chapters (e,g. stub routing) are still completely missing in RFC 7868 \cite{EigrpRFC} even though they are fully operational on Cisco devices. Therefore, we are using network emulator EVE-ng with IOS version \textit{15.7(3)M2} to build our referential topology.

\subsection{Building of Testing Topology}
\begin{figure}[!htp]
    \centering
    \includegraphics[scale=0.6]{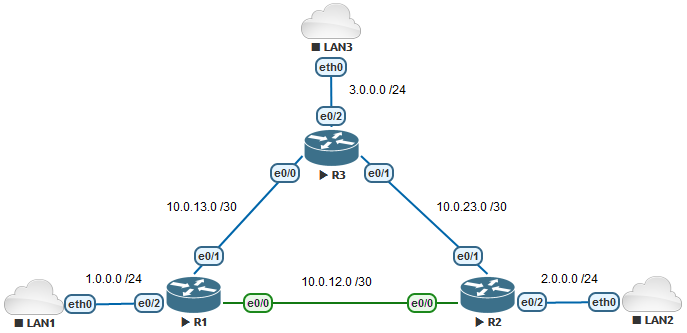}
    \caption{EIGRP testing topology. Scenarios are executed by manipulation of link highlighted in green.}
    \label{fig:topology}
\end{figure}

We decided to use the topology shown in Figure \ref{fig:topology}. It consists of three routers connected to each other via 10 Mbps Ethernet links. EIGRP is configured on all routers and enabled on all interfaces. Additionally, each router has its own LAN which is advertised by the EIGRP process. This demonstration considers two scenarios (with identical configuration).

\begin{itemize}
    \itemsep-0.2em
    \item \textbf{Scenario I: Initial Route Discovery}: The first scenario focuses on the neighbor discovery, adjacency establishment, and initial routing information exchange process. These are typical activities when a new router/link is introduced to the topology. In this scenario, a link between routers \texttt{R1} and \texttt{R2} is added. This link is highlighted green in the Figure \ref{fig:topology}. When this link is introduced, all routers are already configured for EIGRP and \texttt{R1} and \texttt{R3}, as well as \texttt{R3} and \texttt{R2}, are already neighbors and have exchanged routing information. The addition of this link is the starting event for the measurements. Scenario investigates traffic between \texttt{R1} and \texttt{R2} and \texttt{R1}'s routing table.

    \item \textbf{Scenario II: Topology Change Propagation}: The second scenario focuses on route selection and propagation of change in the topology. In this scenario, a link between \texttt{R1} and \texttt{R2}, previously introduced to the topology in Scenario I, is removed from the topology after the topology has reached convergence. The disconnected link is highlighted green in the Figure \ref{fig:topology} and the removal of this link is the starting event for the measurements. This scenario focuses on the traffic between \texttt{R1} and \texttt{R3} and \texttt{R1}'s routing table. The result is expected to be influenced by the race condition as both \texttt{R1} and \texttt{R2} will try to advertise this change in the topology to \texttt{R3} as soon as possible.

\end{itemize}

\subsection{Baseline Production}
Because EVE-ng handles clock synchronization automatically, we just prepared tailored  Cisco EEM applets to execute previously mentioned scenarios on the referential topology. EVE-ng is capable to dump traffic on selected interfaces into \texttt{.pcap} file that can be analyzed. Wireshark view for the captured traffic on the referential topology is shown in Figures \ref{fig:cisco-SC1} and \ref{fig:cisco-SC2}.

We prepared appropriate XML files for \texttt{ScenarioManager}, which manages intended scenarios for a simulation. Traffic generated by simulation models can be viewed in the \textit{Message/packet traffic} window, and each message can be individually inspected. Some simulation models even offer export to PCAP, but this is not true for our EIGRP model as of this moment. Captured traffic in the simulation is shown in Figures \ref{fig:sim-SC1} and \ref{fig:sim-SC2}.

\subsection{Comparison: Protocol Level}
For this particular comparison, we focus on messages' \textit{format}, \textit{content}, \textit{order} and \textit{context}. While format can be checked by a simple comparison of messages of the same type, content, order, and context are harder to compare as it requires a good understanding of the protocol. 

The overall packet length can be misleading (at least for this case). There are genuine implementation differences, e.g., the EIGRP simulation model advertises \texttt{Stub routing information} in every \textit{EIGRP Hello} message while Cisco includes \texttt{Software Version} and \texttt{Peer Topology ID List}. Because aspects like messages' content and order on the referential topology can be affected by the control plane randomness, it is generally beneficial to conduct multiple measurements and closely analyze the one most compatible with the simulation.

\subsubsection{Scenario I}
\begin{figure}[!htp]
    \centering
    \includegraphics{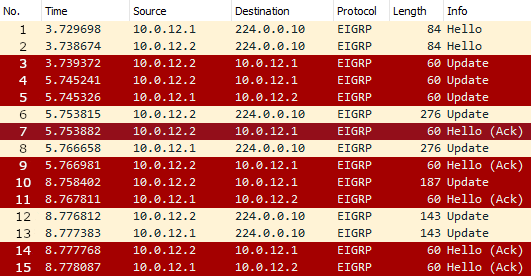}
    \caption{Scenario I - Captured EIGRP traffic between \texttt{R1} and \texttt{R2} displayed with Wireshark}
    \label{fig:cisco-SC1}
\end{figure}

\begin{figure}[!htp]
    \centering
    \includegraphics[scale=0.75]{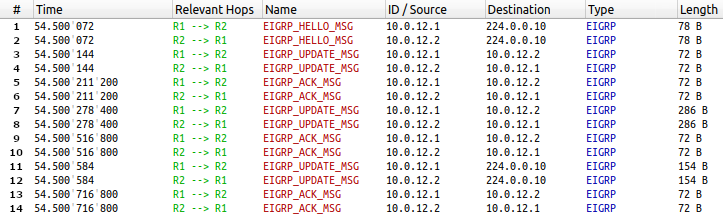}
    \caption{Scenario I - Captured EIGRP traffic between \texttt{R1} and \texttt{R2} diplayed in OMNeT++}
    \label{fig:sim-SC1}
\end{figure}

\begin{figure}[!htp]
    \centering
    \includegraphics[scale=0.7]{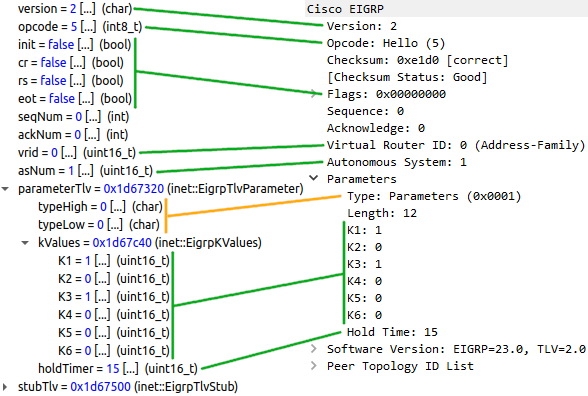}
    \caption{Scenario I - Comparison of Hello messages between referential topology and OMNeT++ simulation.}
    \label{fig:compHello-SC1}
\end{figure}

\begin{figure}[!htp]
    \centering
    \includegraphics[scale=0.7]{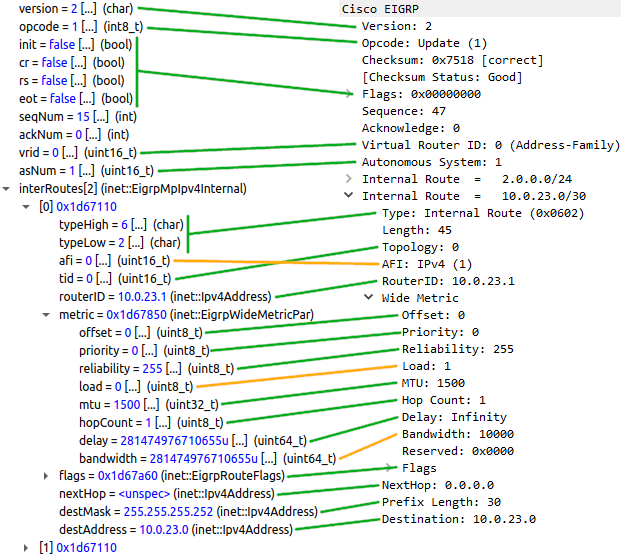}
    \caption{Scenario I - Comparison of Update message \textbf{13} from referential topology and Update message \textbf{11} from the OMNeT++ simulation. }
    \label{fig:compUpdate-SC1}
\end{figure}

Captured traffic between routers \texttt{R1} and \texttt{R2} is shown in Figures \ref{fig:cisco-SC1} and \ref{fig:sim-SC1}. This traffic is compared to each other and most corresponding messages are aligned to the same row in Table \ref{tab:match-sc1}. A closer inspection of messages is shown in Figures \ref{fig:compHello-SC1} and \ref{fig:compUpdate-SC1}.
The referential topology produces very consistent results in this scenario with the race condition causing only marginal differences between measurements. This makes the comparison rather straightforward.

\subsubsection{Scenario II}
\begin{figure}[!htp]
    \centering
    \includegraphics{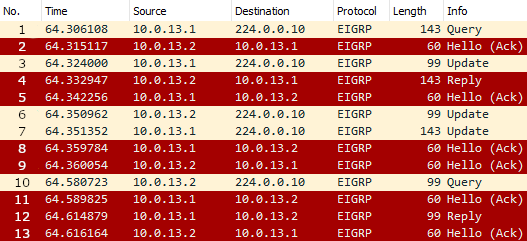}
    \caption{Scenario II - Captured EIGRP traffic between \texttt{R1} and \texttt{R3} displayed with Wireshark}
    \label{fig:cisco-SC2}
\end{figure}

\begin{figure}[!htp]
    \centering
    \includegraphics[scale=0.75]{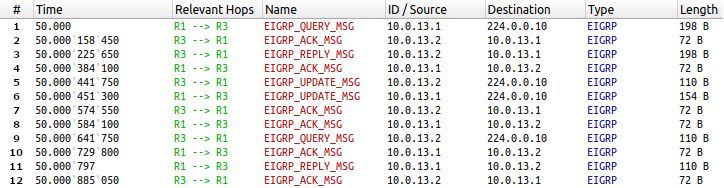}
    \caption{Scenario II - Captured EIGRP traffic between \texttt{R1} and \texttt{R3} diplayed in OMNeT++}
    \label{fig:sim-SC2}
\end{figure}

\begin{figure}[!htp]
    \centering
    \includegraphics[scale=2.2]{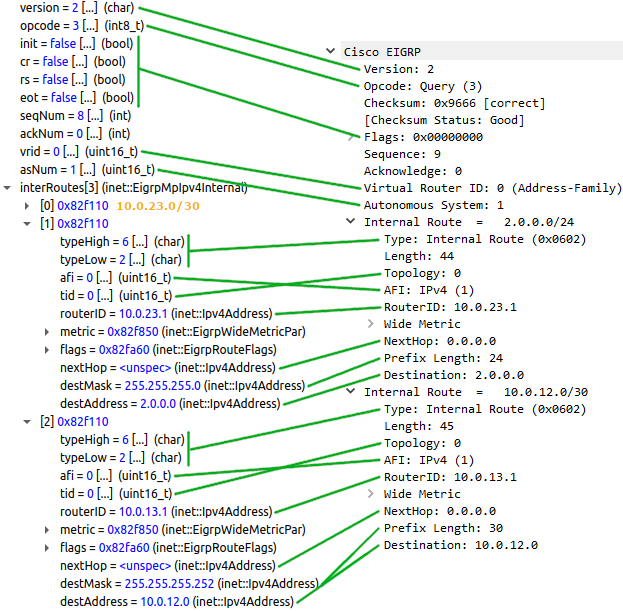}
    \caption{Scenario II - Comparison of Query message \textbf{1}}
    \label{fig:compQuery-SC2}
\end{figure}

Captured traffic between routers \texttt{R1} and \texttt{R3} is shown in Figures \ref{fig:cisco-SC2} and \ref{fig:sim-SC2}. This traffic is compared and most corresponding messages are put on the same row in Table \ref{tab:match-sc2}. A closer inspection of an \textit{EIGRP Query} is shown in Figures \ref{fig:compQuery-SC2}. The result on the referential topology is affected by the race condition which caused the disconnected route \texttt{10.0.12.0/30} to become unreachable on router \texttt{R1} at the very end instead of on the first received \textit{EIGRP Reply}.
\subsubsection{Results}
Even though all messages were contextually correct, the simulation model is not perfect in this regard. The first scenario shows multiple message format errors in TLVs. Another deviation from the referential implementation the model exhibited is the unnecessary use of unicast during the initial synchronization. The second scenario shows an error that causes a route with multiple successors to change into the Active state when one of these successors is lost, which leads to unnecessary traffic. Overall, traffic patterns are easily recognizable and the simulation models still exchanged all of the key information needed for correct changes in data structures like routing or topology table.

\subsection{Comparison: Routing Table Level}
This comparison focuses on changes to the abstract data structures such as routing table, EIGRP topology, or neighbor tables which all can be used to assess the behavior of the simulation model. The snapshot of each of these tables is recorded at the beginning and at the end of the scenario. After the measurement, we perform simple side-by-side comparison while concerning namely: \textit{route source} (identifies how the route was learned, EIGRP labels it with the letter \texttt{D}); \textit{destination} (identifies destination network and mask); \textit{metric} (identifies a value assigned to reach the destination network); \textit{next-hop} (identifies the address of next router); \textit{exit-interface} (identifies the outgoing interface).

Because the configuration is identical between the referential topology and the simulation, all of the above-mentioned fields should be the exact same.
\subsubsection{Scenario I}
\begin{figure}[!htp]
    \centering
    \includegraphics[scale=1.9]{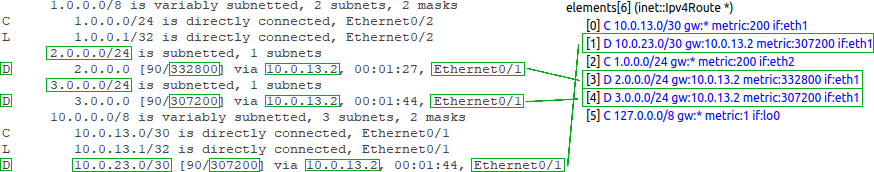}
    \caption{Scenario I - Comparison of router \texttt{R1}'s routing table in its initial state.}
    \label{fig:compRT-SC1}
\end{figure}

Figure \ref{fig:compRT-SC1} depicts the comparison of initial states of the \texttt{R1}'s routing table. Destination \texttt{10.0.12.0/30} is not available at this point in either case. After the new link between routers \texttt{R1} and \texttt{R2} is established, it takes only a few seconds for \texttt{R1} to make necessary changes to the routing table. Figure \ref{fig:compRT-SC1_after} shows the state of \texttt{R1}'s routing table after the topology has reached convergence in both cases.
\subsubsection{Scenario II}
\begin{figure}[!htp]
    \centering
    \includegraphics[scale=1.9]{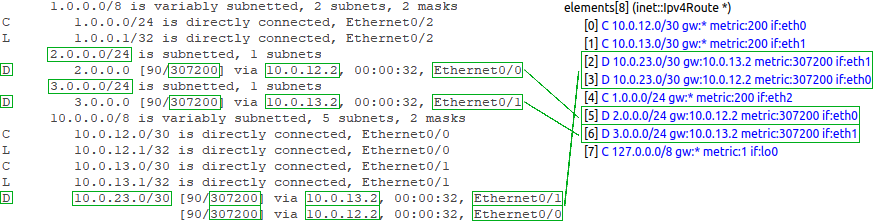}
    \caption{Scenario I - Comparison of router \texttt{R1}'s routing table after the topology reached convergence.}
    \label{fig:compRT-SC1_after}
\end{figure}

\begin{figure}[!htp]
    \centering
    \includegraphics[scale=1.9]{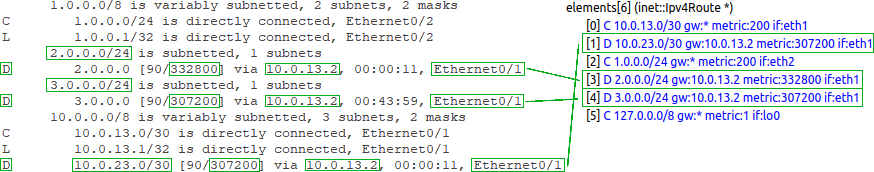}
    \caption{Scenario II - Comparison of router \texttt{R1}'s routing table after the topology reached convergence. Content is identical to the initial state shown in Figure \ref{fig:compRT-SC1}.}
    \label{fig:compRT-SC2}
\end{figure}

The initial state for this scenario is the same as the final state from the first scenario (i.e., \ref{fig:compRT-SC1_after}). The final state of the routing tables for this scenario is shown in Figure \ref{fig:compRT-SC2}. All measured metrics are as expected.
\subsubsection{Results}
In both scenarios, there was no difference in any EIGRP routing table entry between the simulation and the referential configuration. 

\subsection{Revisiting the Implementation}

With this new information about the quality of the model, the programmer is able to proceed with fixing each individual error. While errors like the ones found in Scenario I are easy to fix, as they are mostly just badly predefined numerical values, errors in the FSM like the one found in Scenario II can be much more complex. It is crucial to repeat the previous step in the V\&V process after the changes are introduced to the model.

\subsection{V\&V Reproduction Package}

All necessary files needed to reproduce our experiments are published on \cite{resultrepro}. Results from our measurements are also included.

\section{Conclusion}
\label{sec:conclusion}

In this paper, we have mentioned factors impacting the development of simulation models.
We have outlined challenges and provided lessons learned for the V\&V process, primarily for the domain of computer network protocols.
The main contribution of this paper is the transparent methodology that constitutes several phases aiming at improving the overall quality of the resulting simulation model.
We demonstrated this methodology on the use-case of our recent contribution \cite{pullrequest} to INET, which involves full-fledged support of the EIGRP routing protocol.
We hope this paper will stimulate discussion within the OMNeT++ community (and hopefully beyond it), which would help find a common agreement on the V\&V process for any contributions.

This work was supported by the Brno University of Technology organization and by internal research grant FIT-S-20-6293.

\newpage

\label{sect:bib}
\bibliographystyle{plain}
\bibliography{easychair}

\newpage

\renewcommand\thesubsection{\Alph{subsection}}
\subsection{Configuration Snippets}

\begin{figure}[h!]
    \begin{tcolorbox}[boxsep=-1mm, standard jigsaw, opacityback=0, fontupper=\small, nobeforeafter]
    \texttt{<scenario>}\\
    \texttt{\hspace*{1pc}<at t="50">}\\
    \texttt{\hspace*{2pc}<disconnect src-module="R1" src-gate="ethg[0]"/>}\\
    \texttt{\hspace*{1pc}</at>}\\
    \texttt{\hspace*{1pc}<at t="100">}\\
    \texttt{\hspace*{2pc}<connect src-module="R2" src-gate="ethg[0]" dest-module="R1" \hspace*{2pc} dest-gate="ethg[0]" channel-type="inet.node.ethernet.Eth10M"/>}\\
    \texttt{\hspace*{1pc}</at>}\\
    \texttt{</scenario>}
    \end{tcolorbox}
    \caption{The configuration of ScenarioManager for demonstration from Section \ref{sec:demo} which disconnects and reconnects the link between routers \texttt{R1} and \texttt{R2} at 50 and 100 seconds of simulation time respectively.}
    \label{fig:scenariomanager}
\end{figure}

\begin{figure}[!ht]
    \begin{tcolorbox}[boxsep=-1mm, standard jigsaw, opacityback=0, fontupper=\small, nobeforeafter]
    \texttt{R(config)\# event manager applet SHUTDOWN-APPLET}\\
    \texttt{R(config-applet)\# event timer cron cron-entry "30 15 * * *"}\\
    \texttt{R(config-applet)\# action 1.0 cli command "enable"}\\
    \texttt{R(config-applet)\# action 2.0 cli command "configure terminal"}\\
    \texttt{R(config-applet)\# action 3.0 cli command "interface ethernet 0/0"}\\
    \texttt{R(config-applet)\# action 4.0 cli command "shutdown"}\\
    \texttt{R(config-applet)\# action 5.0 cli command "end"}\\
    \texttt{R(config-applet)\# action 6.0 syslog msg "Applet executed."}
    \end{tcolorbox}
    \caption{Cisco EEM Applet which executes \texttt{shutdown} command on interface \texttt{ethernet 0/0} at a given time according to the router's clock, \textit{15:30} in this case.}
    \label{fig:applet}
\end{figure}

\subsection{Testing topology}

\FloatBarrier

\subsection{Comparison: Scenario I - Initial Route Discovery}

\begin{table}[!htp]
\centering
\renewcommand{\arraystretch}{1,5}
\begin{tabular}{  c : c : p{10cm}  }

Cisco           & OMNeT++   &  Description \\ \hdashline

1, 2             &  1, 2               &   

Exchange of Hello packets. When a router receives a Hello message from a new neighbor, it creates a new entry for this specific neighbor and sets its status to \texttt{pending}. The content and format of these messages are shown in Figure \ref{fig:compHello-SC1}.                 
\vspace{0.1pc}
\\  \hdashline
3, 4, 5            & 3, 4               & 

Exchange of Update packets with \texttt{INIT} flag. These do not contain any routing information. On the referential topology router \texttt{R1} did not acknowledge message \textbf{3} in time, so router \texttt{R2} re-sent the Update as message \textbf{4}, message \textbf{5} contains piggybacked acknowledgement for this message.
\vspace{0.1pc}
\\ \hdashline
6                &          -         & 

This Update message contains advertised routes from router \texttt{R2} and only appears on the referential topology. This message is sent because the neighbor status from \texttt{R2}'s point of view went from \texttt{pending} to \texttt{up}. This causes the message to be ignored and not acknowledged by router \texttt{R1} because from its point of view, \texttt{R2}'s neighbor status is still pending as \texttt{R1} did not receive an acknowledgment for its initial update message, message \textbf{5}, yet.                                 
\vspace{0.1pc}
\\ \hdashline
7                & 5, 6               &

Acknowledgments for initial Update messages. There is only one acknowledgment on the referential topology because it was piggybacked into the Update message as previously mentioned.              
\vspace{0.1pc}
\\ \hdashline
8, 10             & 7, 8               &

Exchange of Update messages containing all advertised routes by both routers. On the referential topology, one Update is sent as unicast because it is a retransmission of message \textbf{6}. It is also smaller because the router applied the \textit{split-horizon} rule which prohibits an advertisement of a route towards its next hop. Another Update on the referential topology is sent as multicast. This is in contrast to the simulation model which uses unicast for the Update messages during the initial synchronization.
\vspace{0.1pc}
\\ \hdashline
9, 11             & 9, 10              &

Acknowledgements for Update messages.
\vspace{0.1pc}
\\ \hdashline
12, 13            & 11, 12             & 

Exchange of Update messages advertising networks which have a successor on this interface as unreachable, i.e, \texttt{2.0.0.0/24} and \texttt{10.0.23.0/30} by \texttt{R1} and  \texttt{1.0.0.0/24} and \texttt{10.0.13.0/30} by \texttt{R2}. This is according to the \textit{poison reverse} rule. The content and format of these messages is shown in Figure \ref{fig:compUpdate-SC1}.                           
\\ \hdashline
14, 15            & 13, 14             & 

Acknowledgments for Update messages.                               
\\ \hdashline
\end{tabular}
\caption{Scenario I - Analysis of the traffic between routers \texttt{R1} and \texttt{R2}.}
\label{tab:match-sc1}
\end{table}

\FloatBarrier

\clearpage
\subsection{Comparison: Scenario II - Topology Change Propagation}

\begin{table}[!htp]
\centering
\renewcommand{\arraystretch}{1,5}
\begin{tabular}{  c : c : p{10,5cm}  }

Cisco           & OMNeT++   &  Description \\ \hdashline

1             &  1             &  

A Query message originating on \texttt{R1} advertising Active states for routes \texttt{10.0.12.0/30} and \texttt{2.0.0.0/24}. Query in the simulation also includes route \texttt{10.0.23.0/30} which signifies an error in the DUAL operation. The content of these Queries is shown in Figure \ref{fig:compQuery-SC2}. 
\vspace{0.1pc}
\\  \hdashline
2             &  2             &  

Acknowledgements for Query messages.
\vspace{0.1pc}
\\  \hdashline
3             &  -             &  

This Update message, originating on \texttt{R1}, advertises route \texttt{10.0.23.0/30} as unreachable. This is according to the \textit{poison reverse} rule and it shows this route being correctly in the Passive state on the referential topology. Because this route is in the Active state in the simulation, router \texttt{R1} has to wait for a Reply which delays this Update until message \textbf{6}. Acknowledgment for this Update is piggybacked in the following Reply.
\vspace{0.1pc}
\\  \hdashline
4             &  3             &  

A Reply for the previous Query, containing \texttt{R3}'s information about given routes. This message contains a metric for route \texttt{2.0.0.0/24} in both cases. Route \texttt{10.0.12.0/30} is also advertised with metric on the referential topology as router \texttt{R3} did not yet receive any Queries from \texttt{R2}. This contrasts with the simulation where router \texttt{R3} advertises \texttt{10.0.12.0/30} as unreachable because it has already received the Query from router \texttt{R2}. A metric for route \texttt{10.0.23.0/30} is also included in the simulation as it was present in the Query.
\vspace{0.1pc}
\\  \hdashline
5             &  4             &  

Acknowledgements for Reply messages.
\vspace{0.1pc}
\\  \hdashline
6             &  5             &  

Update message advertising route \texttt{10.0.12.0/30} with metric on the referential topology and as unreachable in the simulation.
\vspace{0.1pc}
\\  \hdashline
7             &  6             &  

Update message advertising unreachable routes due to the \textit{poison reverse} rule. It contains route \texttt{2.0.0.0/24} in both cases, route \texttt{10.0.12.0/30} is present only on the referential topology and route \texttt{10.0.23.0/30}, equivalent to message \textbf{3} on the referential topology, is present only in the simulation.
\vspace{0.1pc}
\\  \hdashline
8, 9             &  7, 8             &  

Acknowledgements for Update messages.
\vspace{0.1pc}
\\  \hdashline
10             &  9             &  

Query message advertising route \texttt{10.0.12.0/30} in the Active state for router \texttt{R3}. This message is caused by the arrival of a Query from router \texttt{R2}.
\vspace{0.1pc}
\\  \hdashline
11             &  10             &  

Acknowledgements for Query messages.
\vspace{0.1pc}
\\  \hdashline
12             &  11             &  

Reply advertising \texttt{10.0.12.0/30} as unreachable.
\vspace{0.1pc}
\\  \hdashline
13             &  12             &  

Acknowledgements for Reply messages..
\vspace{0.1pc}
\\  \hdashline

\end{tabular}
\caption{Scenario II - Analysis of the traffic between routers \texttt{R1} and \texttt{R3}}
\label{tab:match-sc2}
\end{table}


\end{document}